\documentclass[aps,prb,twocolumn,groupedaddress]{revtex4}
\usepackage[english]{babel}
\usepackage{graphicx}

\begin{document}

\title{Disorder-based graphene spintronics }
\author{A. R. Rocha$^1$}
\author{Thiago B. Martins$^2$}
\author{A. Fazzio$^{1,2}$}
\author{Ant\^onio J. R. da Silva$^2$}
\email{ajrsilva@if.usp.br}
\affiliation{$^1$Centro de Ci\^ encias Naturais e Humanas, Universidade Federal do ABC, Santo Andr\' e, SP, Brazil}
\affiliation{$^2$Instituto de F\'{\i}sica, Universidade de S\~ao Paulo, CP 66318, 05315-970, S\~ao Paulo, SP, Brazil}

\begin{abstract}
 The use of the spin of the electron as the ultimate logic bit - in what has been dubbed spintronics - can lead to a novel way of thinking about information flow. At the same time single layer graphene has been the subject of intense research due to both its fundamental properties as well as its potential application in nanoscale electronics. While defects can significantly alter the electronic properties of nanoscopic systems, the lack of control can lead to seemingly deleterious effects arising from the random arrangement of such impurities. Here we demonstrate, using {\it ab initio} density functional theory and non-equilibrium Green's functions calculations, that it is possible to obtain perfect spin selectivity in doped graphene nanoribbons to produce a perfect spin filter. We show that initially unpolarized electrons entering the system give rise to 100\% polarization of the current due to random disorder. This effect is explained in terms of different localization lengths for each spin channel which together with the well know exponential dependence of the conductance on the length of the device leads to a new mechanism for the spin filtering effect that is enhanced by disorder.
\end{abstract}

\maketitle

The possibility of using the spin of the electron - instead of its charge - as information carrier opens tantalizing possibilities. In what has come to be known as spintronics the spin would provide us with the ultimate logic bit.\cite{wolf_science} In that context, carbon-rich materials have appealing prospects for spin-dependent transport due to small spin-orbit and hyperfine interactions; two of the main sources of spin dephasing. Spintronics entered the realm of carbon-based materials with the seminal work of Tsukagoshi {\it et al.}\cite{tsukagoshinature_cnt} who have shown that the spin coherence length of electrons injected from ferromagnetic electrodes into carbon nanotubes  has a lower limit of 130 nm. Presently one of the most exciting amongst graphitic materials is graphene.\cite{geim_science} Just recently single layer graphene has been isolated and some of its unique properties measured.\cite{geim_review_nat,geim_nature,phillipkim_nat,castroneto_rmp} In particular, some of the results for graphene worth citing in the scope of this work are the  experimental observation that the spin coherence length in graphene can reach the micrometer range,\cite{wees_nature} theoretical predictions of giant magnetoresistance effects in graphene nanoribbons (GNRB)\cite{kim_natmat} and finally the use of an external electric field to tune the possible half-metalicity of the ribbons.\cite{louie_nature}

Until now, most works proposing nanoscopic spintronics devices have dealt with spin injection from magnetic electrodes into a non-magnetic medium using a prototypical magnetoresistance setup.\cite{baibich_gmr,binasch_gmr} In this work we propose a fundamentally different approach, where the device itself acts as the active region  - not the electrodes. Moreover, both the device itself and the electrodes are built in the same material framework. This way one is able to select the spin of the electrons flowing through the system without the need to resort to heterojunctions or multilayered materials.

The electronic properties of systems such as carbon nanotubes and graphene can be altered by the introduction of impurities, enhancing even more their potential for applications.\cite{cnxsensors,rochaprl,graphsensors} A possible candidate that presents a combination of both low formation energy and significant changes to the transport properties of GNRBs is substitutional Boron.\cite{thiagoprl,thiagonanolett} Boron in carbon nanotubes and graphite have already been reported\cite{cntdoping,ruhle_cpl,bdoping_graphite_jap} and there are suggestions of small spin polarization in singly doped GNRBs.\cite{thiagoprl}

There are other proposals of spin-filtering effects in the literature related mostly to the nanoribbon edge termination ranging from different end-groups\cite{katsnelson} to different orientations in the same GNRB.\cite{tomanek} However none of which present 100\% spin-filtering. Moreover - and most importantly - although significant level of control has been achieved in creating ever smoother lateral edges via chemical processes,\cite{daiscience2008} it is highly unlikely that one will be able to assemble a device with a single impurity atom - or a few impurities placed in carefully chosen positions - at an industrial scale. As it is usually the case, the presence of disorder would seemingly have a detrimental effect over the degree of polarization of the system. Therefore one raises the fundamental question of whether conductance polarization would survive in the presence of a large number of defects randomly distributed along a one-dimensional system up to 0.5 $\mu m$ in length in what would be considered a more realistic setup.

Although computer simulations can help determine the behavior of nanoscopic devices, up until now atomistic {\it ab initio} predictions that can be directly related to experiments in this length scale had been hindered by the shear size of the calculation - up to 40,000 atoms - required to answer the question posed above. In this work,
combining {\it ab initio} density functional theory (DFT)\cite{dft,dft2} with recursive non-equilibrium Green's functions (NEGF)\cite{rochaprl,sanvito,brandbyge} techniques for calculating electronic transport, we propose a true spin filter based on realistic B-doped graphene nanoribbons with up to 450 nm in length. We present results for Boron as a point in case in what is a much more general effect. We show that in disordered systems the current can reach 100\% spin polarization even if the electrons entering the device are not spin polarized at all, provided there is a spin-asymmetry in the single scatterer transmission probabilities. Furthermore, we demonstrate that disorder has an important role in the enhancement of this spin selectivity effect. In fact, it becomes evident that different localization lengths for each spin channel combined with the exponential behavior of the conductance in the Anderson localization\cite{anderson,kramer} regime is what gives rise to this remarkable result.

\begin{figure}[h]
\center
\includegraphics[width=8.0cm,clip=true]{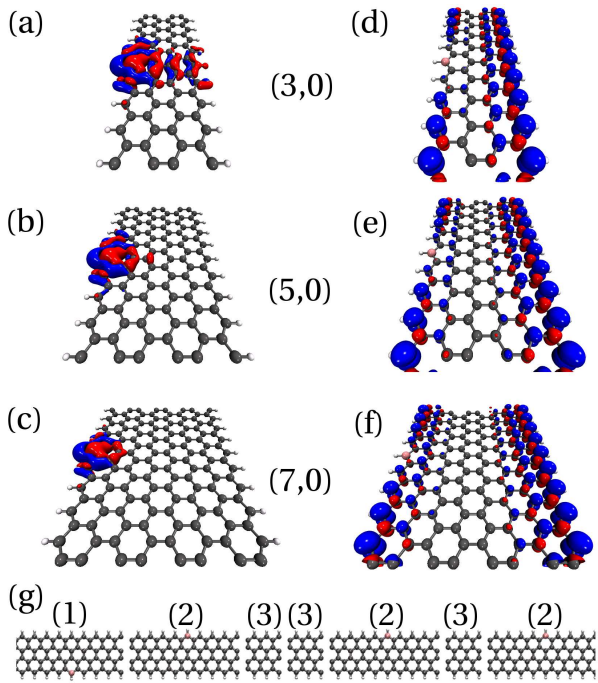}
\caption{Difference in charge density upon introduction of a Boron atom compared to pristine carbon nanoribbons of different widths: a) (3,0), b) (5,0), and c) (7,0). d-f) Magnetic moments for the same nanoribbons with different widths. In all cases blue (red) represent positive (negative) values of the iso-surface fixed at +(-) $2\times 10^{-3}~ e/Bohr^3$. g) Segment of a disordered GNRB. Each of the three types of building blocks used in our calculations are shown. Color code: H - white, C - grey, and B - light red.}\label{chargemagmoment}
\end{figure}

Initially we performed DFT calculations on graphene nanoribbons of different widths with one substitutional Boron atom. In all cases 9 irreducible cells for a Graphene nanoribbon were used. The calculations were performed using the SIESTA code.\cite{siesta2} Our calculations show that the formation energy of substitutional Boron at the edges of the nanoribbon is smaller than any other site by at least $0.8~eV$.\cite{thiagoprl,huang} Furthermore, our calculations also indicate that B atoms do not clusterize to the extent that two Borons prefer to stay as far apart as possible as to remain non-interacting. Hence in all  calculations presented hereafter we assume that the B atoms are positioned  at either edges of the ribbon. The ground state of a zig-zag pristine carbon nanoribbon presents ferromagnetic coupling between atoms of the same edge whereas the ordering is anti-ferromagnetic between the two edges. It has been noted, however, that the energy differences between the lowest energy solution and the fully ferromagnetically oriented GNRB is relatively small (approximately 11 meV per edge atom) in which case at room temperature the edges are decoupled and a small magnetic field could be applied simply to align the magnetic moments of the pristine nanoribbon without significantly affecting the electronic structure of the system. In all calculations shown here we use the fully ferromagnetic solution since it gives rise to a metallic state and would keep the magnetic moments fully aligned even at room temperature.\cite{katsnelson1}

\begin{figure}[h]
\includegraphics[width=8.0cm,clip=true]{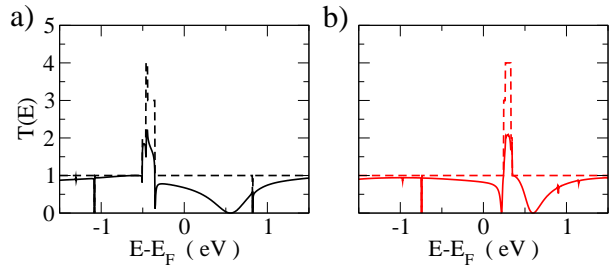}
\caption{a-b) Total transmission coefficients as a function of energy for a (3,0) carbon nanoribbon with a single boron impurity. The dashed lines indicate the transmission coefficients of a pristine GNRB. Black (red) curves indicate majority (minority) spin channels.}\label{transmission}
\end{figure}

From our results we look at the changes in the electronic structure of nanoribbons caused by the presence of Boron impurities. In figure \ref{chargemagmoment}(a-c) we show the difference in charge density $\Delta\rho \left({\mathbf r}\right)$ calculated with DFT between a pristine nanoribbon and one with substitutional Boron. We note that, for a narrow GNRBs the introduction of the impurity leads to changes in the charge density that are localized along the device, but that spill onto the border opposite to the defect. As we make the system wider - going from a (3,0) to a (5,0) and finally to a (7,0) GNRB - the changes in charge density quickly localize around the impurity. Finally one can clearly see that the magnetic moment ($\rho^\uparrow\left({\mathbf r}\right) - \rho^\downarrow\left({\mathbf r}\right)$) at the edges close to the B atom goes to zero (figure \ref{chargemagmoment}(d-f)) due to the empty $p_z$ orbital for both spins.

We initially consider one segment of a (3,0) GNRB containing a single Boron
impurity - as seen in figure \ref{chargemagmoment}a - coupled to two semi-infinite pristine nanoribbons to either side acting as charge reservoirs. We use standard NEGF formalism coupled to DFT.\cite{rocha,frederico} Figure \ref{transmission} shows the transmission for majority and minority spins for both a pristine GNRB and one with a single Boron impurity. One notes that total transmission results for the single impurity case at the Fermi level, $E_F$ are 0.67 and 0.82 for majority and minority spins respectively. In this work we are interested in the spin filtering effects. One important measure of a device's capability to act as a spin filter is the degree of polarization $P$ defined here as $ P = \left(g^\uparrow-g^\downarrow\right)/\left(g^\uparrow+g^\downarrow\right)$,
where $g^\sigma$ is the spin dependent ($\sigma=\left\{\uparrow,\downarrow\right\}$) total conductance $G^\sigma$ normalized by $e^2/h$. Consequently, the conductance polarization for the single impurity case is approximately -10\%, similar to what has been reported in earlier works.\cite{thiagonanolett}

One important point that must be noted here is that, although one might say that the pristine system is ferromagnetic, at $E_F$ the polarization of the current is exactly zero. Hence, deep inside the leads the current is not spin-polarized at all, and the total transmission is equal to 1 for both spin channels thus justifying our choice of electrodes.

\begin{figure}[h]
\includegraphics[width=8.0cm,clip=true]{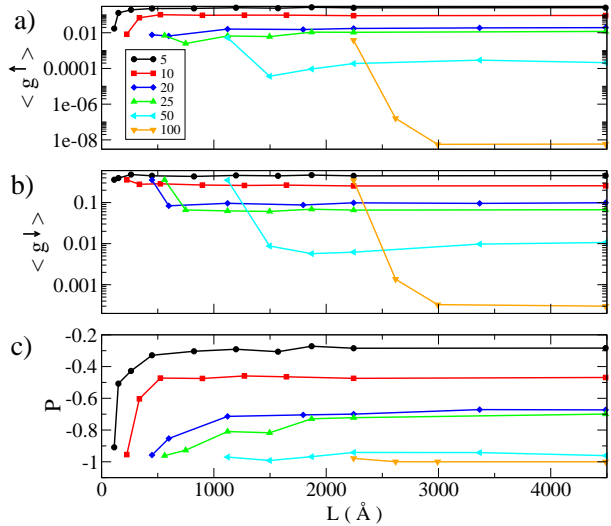}
\caption{Normalized conductance as a function of length for (3,0) graphene nanoribbons with a different number of scattering centers: a) majority and b) minority spin channels. c) Degree of polarization.}\label{fig03}
\end{figure}

We then set out to simulate a more realistic device where Boron is incorporated at random along the GNRBs. In order to do so we use a combination of DFT and recursive Green's function methods.\cite{rochaprl,sanvito,brandbyge} Our device is modeled by dividing a long system into segments (as shown in figure \ref{chargemagmoment}g) consisting of regions with impurities - in this case the two different edges of the nanoribbon where the B atom can be positioned - and  pieces of the pristine graphene nanoribbon. A segment of the GNRB highlighting the building blocks (indexes $1$, $2$ and $3$) of our device is depicted in figure \ref{chargemagmoment}g. We can clearly see that the distance between two B atoms may vary as well as the edge where it is placed.

In our calculations we have considered different defect concentrations per mass (obtained by changing the number of scattering sites containing Boron) ranging from approximately $0.9\%$ down to $0.3\%$; the length of the ribobns were varied up to $450~nm$ and the temperature ranged from 300 down to 3 K.

In figure \ref{fig03}(a-b) we present spin-resolved conductance calculations  as a function of the length $L$ of the nanoribbon. Each curve represents a different number of scattering sites $N$ (the defect concentration for each curve decreases as we increase the length). Two important aspects can be noticed from this curve. The first one is that, after a characteristic length the conductance for both majority and minority spins saturates. The saturation of the conductance can be understood in terms of Anderson localization. In this regime it has been shown that, besides the well known exponential dependence of the conductance on the length of the device, $G \propto e^{-L/\xi}$,\cite{kramer} the localization length $\xi$ is inversely proportional to the concentration of scatterers, $\rho = N/L$ along the nanoribbon.\cite{flores2,roche_blase_rmp_2007} Hence, simple algebra shows that, for a fixed number of scattering centers, one would expect the conductance to be independent of the length of the system when we are in the localized regime.

The second feature however is also the most striking one. From figure \ref{fig03}c we note that the degree of spin polarization, $P$ increases with the number of defects until the current becomes essentially 100\% spin polarized when the number of scattering centers is of the order of 100 for a range of concentrations and GNRB lengths. Thus, remarkably, the presence of random disorder has not only preserved the conductance polarization, but in reality has led to an enhancement of the spin filtering effect until majority spins are completely blocked.

\begin{figure*}[t]
\includegraphics[width=15cm,clip=true]{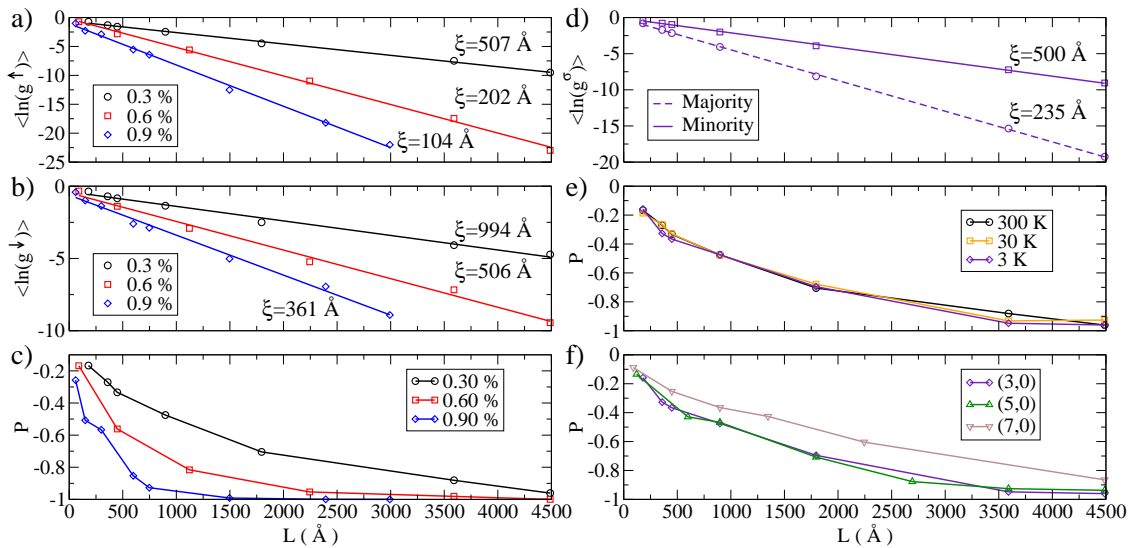}
\caption{Average of $\ln \left(g^{\sigma}\right)$ of a (3,0) GNRB as a function of the length of the device for a) $\sigma=\uparrow$ and b) $\sigma=\downarrow$ in cases of different concentration and room temperature. c) Conductance polarization at room temperature. d) Average of  $\ln \left(g^{\sigma}\right)$ for T=3K and 0.3\% concentration. e) Conductance polarization at different temperatures as a function of GNRB length for impurity concentration fixed at 0.3\%. f) Conductance polarization for GNRB of different widths. The concentration was kept at 0.3\%, and T=3~K.}\label{fig04}
\end{figure*}

This, in principle, seems counter-intuitive. In order to understand this behavior we look into the characteristic length of our device, namely the localization length. In figure \ref{fig04} we show the average of the natural logarithm of the normalized conductance for each spin channel as a function of system length. For all defect concentrations shown we can clearly see that the points fall nicely into a straight line as predicted by Anderson for the localized regime.\cite{anderson,kramer} The main point here is that one obtains different localization lengths for each spin channel in such a manner that the ratio $\xi^\uparrow/\xi^\downarrow$ ranges between 0.51 - 0.29 depending on the concentration of defects. With the ratio between the minority and majority localization lengths being approximately two, this means that $g_{\uparrow}\sim g_{\downarrow}^2$. In other words, an increase in the length of the device such that the spin down conductance decreases by a factor of ten implies that the spin up conductance decreases much faster, by a factor of one hundred.

The mechanism for this spin filtering effect can be traced back to scaling theory of localization.\cite{brandbyge_prl_2007,roche_blase_mplb_2007} There one can relate the conductance at a certain impurity concentration to the resistance of a single scattering site and to the mean free path, $l_e$. In turn, the mean free path can be directly associated, using Thouless relation,\cite{roche_blase_mplb_2007,beenakker} to the localization length. For our particular case of a GNRB one would find that, at a fixed concentration and zero temperature, the ratio between the spin-dependent localization lengths would be equal to the inverse ratio between the single impurity resistances for each spin channel, $\xi^\uparrow/\xi^\downarrow=R_s^\downarrow/R_s^\uparrow$. For a single impurity in the specific case of a GNRB the resistance can be approximated by $R_s^\sigma=h/e^2\left(1/T^\sigma \left(E_F \right)-1\right)$, where one has to remove the contact resistance $2/G_0$ for each spin.
From figure \ref{transmission} we can estimate the value of $R_s^\downarrow/R_s^\uparrow$ as 0.45. Since this relation is only valid at T=0 we performed calculations at low temperatures (T = 3 K). The results presented in figure \ref{fig04}d show that the localization length changes considerably as we lower the temperature although the curves retain the straight-line character that characterizes the localization regime.\cite{flores2} In the low temperature regime we obtain a value of $\xi^\uparrow/\xi^\downarrow=0.45$. This is in excellent agreement with our predictions for the single impurity case, thus unequivocally demonstrating that this effect is indeed disorder-driven, but most importantly that it can be inferred from single impurity scattering.

Strikingly, from figure \ref{fig04}e, we also observe that the conductance polarization is resilient to changes in temperature even if $\xi^\sigma$ changes considerably. Finally one might argue that (3,0) GNRBs are quite narrow and the effect seen in figure \ref{chargemagmoment}a - where the charge density differences due to the Boron impurity on one edge of the GNRB spills onto the opposite border - could have significant influence over our conclusions. For (5,0) and (7,0)  nanoribbons which are already within state of the art experimental range,\cite{daiscience2008} changes to the density (as seen in figure \ref{chargemagmoment}(b-c)) are confined to only one side. In fact, it has been shown that the degree of polarization of a single B impurity significantly decreases as a function of GNRB width.\cite{thiagoprl} We observe from figure \ref{fig04}f, however, that for the same concentration of defects  the degree of polarization is still very high.

As a final remark the resistance measurements required to obtain a polarization close to $100\%$ would be of the order of 100 M$\Omega$, we can see that one could easily obtain values of $P$ between $60-80\%$ already in the $100~k\Omega$ regime with approximately 25 impurities. This range could be easily achieved by today's experimental apparatus.\cite{wees_nature}

In conclusion, we have proposed a device that presents perfect spin-filtering capabilities. It is based on B-doped graphene nanoribbons that not only retain conductance polarization in the presence of defects, but we observe a staggering increase reaching up to 100\% spin filtering. This effect was explained in terms of different scattering probabilities at the impurity site for majority and minority spins which consequently leads to different localization lengths. The exponential dependence of the conductance on the localization length thus results in changes in resistance that are significant. Therefore, we have proposed a device based on the realistic assumption that impurities are present in an uncontrolled manner. These impurities in fact drive the spin selectivity of the system with the added bonus that one does not need to rely on a spin valve based on multilayered materials and interface problems can be largely ignored. It is important to note that no assumptions about the specific nature of the defect were necessary in our scaling model for the localization length. Thus, although we have shown a proof-of-concept using Boron scatterers one would expect a similar behavior with different scattering centers provided the spin degeneracy of the scattering probability is broken. Finally the results shown here might have important technological implications since it might lead to a new family of electronic devices where one can select the spin of the electron and consequently make a realistic spintronics device.

\section{Materials and methods}
Our density functional theory calculations were performed on graphene nanoribbons of different widths, namely (3,0), (5,0) and (7,0) GNRBs. In all cases 9 irreducible cells for a GNRB were used (see figure \ref{chargemagmoment}). The calculations were performed using the SIESTA\cite{siesta2} code which uses a pseudo-atomic localized basis set. We employed a double zeta basis set with polarization orbitals for all atoms with the Perdew-Burke-Ernzerhof (PBE) parametrization within the Generalized Gradient Approximation.\cite{pbe} Upon substitution of one carbon atom by a boron atom the system is allowed to fully relax by means of a conjugate gradient method.

For our transport calculations we model our device as a central scattering region coupled to two electrodes
(left and right). In the absence of spin-orbit interactions we can consider majority and minority spins as two completely independent systems which have their own transport properties within the two spin fluid approximation.\cite{mott_prsa} The key quantity in our calculations is the Green's function (at an energy $E$) for the central device attached to the two charge reservoirs\cite{sanvito,rocha,frederico}
\begin{equation}\label{green}
{\cal G}^\sigma\left(E\right) = \left[ E \times S - H^\sigma_\mathrm{S} - \Sigma^\sigma_\mathrm{L} - \Sigma^\sigma_\mathrm{R} \right]^{-1} ~,
\end{equation}
where $H_{\mathrm S}^\sigma$ ($\sigma\equiv\left\{\uparrow,\downarrow\right\}$) and $S_\mathrm{S}$ are the Kohn-Sham\cite{dft2} Hamiltonian and overlap matrices for the central scattering region coming from DFT total energy calculations, and $\Sigma^\sigma_\mathrm{L}$ and $\Sigma^\sigma_\mathrm{R}$ are the so-called self-energies which introduce the effects of the electrodes onto the scattering region.\cite{rocha,frederico} The standard Landauer-B\"uttiker formula is then used to calculate the transmission coefficients $T^\sigma\left(E\right)$ ($\sigma=\left\{\uparrow,\downarrow\right\}$).\cite{landauer,caroli}

In order to calculate the electronic transport properties of a more realistic device containing up to 40,000 atoms one must find an efficient way of obtaining $T^\sigma\left(E\right)$ without direct invesion of equation \ref{green}.
One can use the finite-range character of the coupling within the system, and the fact that only a sub-set of the Green's function is actually required to calculate the conductance.  For that purpose the scattering region is split into segments. Each segment - or building block - is obtained from a separate DFT calculation and is constituted of regions with or without defects. Thus, for a given length and concentration, any device can be built by randomly arranging the different building blocks. We then use a recursive Green's function method\cite{sanvito,brandbyge,rocha} that maps the Hamiltonian in equation \ref{green} onto two renormalized Hamiltonians for the two electrodes connected via an effective coupling. Subsequently one can calculate the transmission coefficients using a Landauer-like formula\cite{caroli,landauer}
\begin{equation}
T^\sigma\left(E\right) = Tr\left[\Gamma^\sigma_\mathrm{L}\left(E\right){\cal G}^\sigma_{LR}\left(E\right)\Gamma^\sigma_\mathrm{R}\left(E\right){\cal G}_{LR}^{\sigma\dagger}\left(E\right)\right] ~,
\end{equation}
where ${\cal G}_{LR}^\sigma$ is the block of the Green's function that couples the left and right electrodes and $\Gamma_{\alpha}^\sigma=i\left[\Sigma^\sigma_\alpha-\Sigma^{\sigma\dagger}s_\alpha\right]$.

The differential conductance of the system in the linear regime is then given by
\begin{equation}
G^\sigma=\lim_{V\rightarrow 0} \frac{dI^\sigma}{dV} \sim \frac{e^2}{h}\int T^\sigma\left(E\right)\left.\frac{df\left(E^\prime,E_F\right)}{dE^\prime}\right|_{E} dE~,
\end{equation}
where $f\left(E^\prime,E_F\right)$ is the Fermi distribution with $E_F$ the Fermi level of the electrodes. Temperature effects are taken into consideration within the Fermi distribution. Finally due to the stochastic nature of our calculations up to 500 configurations were randomly generated for each defect concentration and the average was taken.




\end{document}